\documentclass[11pt]{article}

\usepackage{fullpage}
\usepackage{latexsym}
\usepackage{amsfonts}
\usepackage{amssymb}
\usepackage{amsmath}
\usepackage{graphicx, comment}

\usepackage{color}

\makeatletter

\let\c@table\c@figure
\makeatother

\definecolor{suRed}{RGB}{197,0,36}

\newcommand{\oo}{\infty}

\renewcommand{\bar}{\overline}

\newtheorem{thm}{Theorem}[section]

\newcommand{\beq}{\begin{equation}}
\newcommand{\eeq}{\end{equation}}
\newcommand{\ba}{\begin{array}}
\newcommand{\ea}{\end{array}}
\newcommand{\bea}{\begin{eqnarray*}}
\newcommand{\eea}{\end{eqnarray*}}
\newcommand{\bc}{\begin{center}}
\newcommand{\ec}{\end{center}}
\newcommand{\bt}{\begin{table}}
\newcommand{\et}{\end{table}}

\newcommand{\sgn}{\mbox{sgn}}
\newcommand{\la}[1]{\label{#1}}

\newcommand{\ds}{\displaystyle}
\newcommand{\no}{\noindent}
\newcommand{\pp}[2]{{\partial #1 \over \partial #2}}
\newcommand{\ppn}[3]{{\partial^{#1} #2 \over \partial #3^{#1}}}

\newcommand{\rf}[1]{(\ref{#1})}
\newcommand{\beqno}{\begin{displaymath}}
\newcommand{\eeqno}{\end{displaymath}}

\newcommand{\been}{\begin{enumerate}}
\newcommand{\een}{\end{enumerate}}

\newlength{\myheight}
\newlength{\mylength}

\newcounter{saveeqn}

\def\XXint#1#2#3{{\setbox0=\hbox{$#1{#2#3}{\int}$}
     \vcenter{\hbox{$#2#3$}}\kern-.5\wd0}}

\allowdisplaybreaks[1]

\begin{document}

\title{Recovering the water-wave profile from pressure measurements}

\author{K.\ L.\ Oliveras\thanks{oliverak@seattleu.edu}\,, V.\ Vasan\thanks{vvasan@amath.washington.edu}\,, B. Deconinck\thanks{bernard@amath.washington.edu}\,, D.\ Henderson$^\S$\thanks{dmh@math.psu.edu}\\
~\\
$^{*}$Department of Mathematics, Seattle University, Seattle, WA 98122\\
$^{\dag}$ $^{\ddag}$ Department of Applied Mathematics, University of Washington, Seattle, WA 98195-2420\\
$^{\S}$Department of Mathematics, Penn State University, University Park, PA 16802
}

\date{\today}

\maketitle

\begin{center}
{\em This paper is dedicated to the memory of Joe Hammack,\\ who provided the impetus for this project.}
\end{center}

\begin{abstract}
A new method is proposed to recover the water-wave surface elevation from pressure data obtained at the bottom of the fluid. The new method requires the numerical solution of a nonlocal nonlinear equation relating the pressure and the surface elevation which is obtained from the Euler formulation of the water-wave problem without approximation. From this new equation, a variety of different asymptotic formulas are derived. The nonlocal equation and the asymptotic formulas are compared with both numerical data and physical experiments. The solvability properties of the nonlocal equation are rigorously analyzed using the Implicit Function Theorem.
\end{abstract}

\section{Introduction}

In field experiments, the elevation of a surface water-wave in shallow water is often determined by measuring the pressure along the bottom of the fluid, see {\em e.g.} \cite{baquerizo}, \cite{donelan}, \cite{kuo1}, \cite{kuo2}, \cite{tsai1}, \cite{tsai2}. A variety of approaches are used for this. The two most commonly used are the hydrostatic approximation and the transfer function approach. For the hydrostatic approximation \cite{deandalrymple, kundu},

\beq\la{archie}
\eta(x,t)=\frac{P(x,-h,t)}{\rho g}-h,
\eeq

\no where $g$ is the acceleration due to gravity, $h$ represents the average depth of the fluid, $\rho$ is the fluid density, $P(x,-h,t)$ is the pressure as a function of space $x$ and time $t$ evaluated at the bottom of the fluid $z=-h$, and $\eta(x,t)$ is the zero-average surface elevation. Throughout, we assume that all wave motion is one-dimensional with only one horizontal spatial variable $x$. The hydrostatic approximation is used, for instance, in open-ocean buoys employed for tsunami detection, see~\cite{buoy}.

The transfer function approach uses a linear relationship between
the Fourier transforms $\mathcal{F}$ of the dynamical part of the pressure and the elevation of the surface~\cite{deandalrymple, escherschlurmann, kennedy, kundu}:

\begin{equation}\label{engr}
\mathcal{F}\left\lbrace \eta(x,t)\right\rbrace(k)=\cosh(kh)\mathcal{F} \left\lbrace p(x,t)/g \right\rbrace (k),
\end{equation}

\no where $p(x,t)=(P(x,-h,t)-\rho g h)/\rho$ is the dynamic (or non-static) part of the pressure $P(x,z,t)$ evaluated at the bottom of the fluid $z=-h$, scaled by the fluid density $\rho$. In this relationship, $\eta$ and $p$ are regarded as functions of the spatial coordinate $x$, with parametric dependence on time $t$.  It is equally useful to let $t$ vary for fixed $x$, as would be appropriate for a time series measurement, which results in extra factors of the wave speed $c(k)$, due to the presence of a temporal instead of a spatial Fourier transform.

It is well understood that nonlinear effects play a significant role when reconstructing the surface elevation for shallow-water waves or for waves in the surf zone (see \cite{Bergan, donelan, tsai2}, for instance). Since nonlinear effects are not captured by the linear transfer function \rf{engr}, different modifications of \rf{engr} have been proposed. One approach is to modify the transfer function to incorporate extra parameters ({\em e.g.}, multiplicative factors, width scalings) that are tuned to fit data \cite{kuo1, kuo2, tsai2}. A less empirical approach is followed in \cite{Bergan} and \cite{LeeWang} where corrections to the transfer function are proposed based on higher-order Stokes expansions.
Bishop \& Donelan \cite{donelan} examine the empirical approaches and argue that the inclusion of the proposed parameters is not necessary as errors from inaccuracy in instrumentation and analysis are likely to outweigh the benefit of their presence. We do not include any of the modified transfer function approaches in our comparisons.

Bishop and Donelan \cite{donelan} acknowledge that the linear response cannot accurately capture nonlinear effects. While both the hydrostatic model and the transfer function approach are accurate on some scales, they fail to reconstruct the surface elevation accurately in the case of large-amplitude waves, as might be expected. Errors of 15\% or more are common, as is shown and discussed below. In order to address the inaccuracies of the linear models, nonlinear methods are required.  With the exception of recent work by Escher \& Schlurmann \cite{escherschlurmann} and Constantin \& Strauss \cite{constantinstrauss}, few nonlinear results are found in the literature. Escher \& Schlurmann \cite{escherschlurmann} provide a consistent derivation of \rf{engr} and offer some thoughts about the impact of nonlinear effects. Starting from a traveling wave assumption, Constantin \& Strauss \cite{constantinstrauss} obtain different properties and bounds relating the  pressure and surface elevation. However, they do not present a reconstruction method to accurately determine one function in terms of the other.

One way to obtain an improved pressure-to-surface elevation map is to use perturbation methods to determine nonlinear correction terms to (\ref{engr}). Several such approaches are given below, and we include them when comparing the different methods. Our main focus, however, is the presentation of a new nonlocal nonlinear relationship between the pressure at the bottom of the fluid, and the elevation of a traveling-wave surface that captures the full nonlinearity of Euler's Equations. The advantage of this approach is that

\begin{enumerate}

\item it allows for the surface to be reconstructed numerically from any given pressure data for a traveling wave,

\item it provides an environment for the direct analysis of the relationship between all physically relevant
    parameters such as depth and wave speed,

\item and it allows for the quick derivation of perturbation expansions such as the ones mentioned above.

\item Although our approach is formally limited to traveling waves, it can be applied with great success to more general wave profiles that are not merely traveling. This is illustrated and discussed below.

\end{enumerate}

\begin{comment}
In the case of periodic boundary conditions with period $L$ equated to $2\pi$, the relationship is

\begin{equation}\label{periodicawesome}
\frac{\sqrt{c^2 - 2g\eta}}{\sqrt{1 + \eta_x^2}}  = \sum_{k=-\infty}^\infty e^{ikx}\hat{P}_k\cosh\left(k\left(\eta +h\right)\right),
\end{equation}
where
\begin{itemize}
\item  $\hat{P}_k = \frac{1}{2\pi}\int_0^{2\pi} e^{-ikx}\sqrt{c^2-2p(x)}~dx$, and
\item $c$ is the speed of the traveling wave.
\end{itemize}

Similarly, the corresponding equation on the whole line (with sufficiently fast decaying boundary conditions as $x \to \pm \infty$) is
\begin{equation}\label{awesome}
\frac{\sqrt{c^2 - 2g\eta}}{\sqrt{1 + \eta_x^2}}  =  \int_{-\infty}^\infty e^{ikx}\hat{P}(k)\cosh\left(k\left(\eta +h\right)\right)~dk,
\end{equation}
where $\hat{P}(k) = \frac{1}{2\pi}\int_{-\infty}^{\infty} e^{-ikx}\sqrt{c^2-2p(x)}~dx$.
\end{comment}

In what follows, we derive these nonlocal relations and demonstrate their practicality. We compare results from the nonlocal formulation with those from the linear approaches and different nonlinear perturbative models, using both numerical data for traveling waves in shallow water, and experimental data obtained at Penn State's Pritchard Fluid Mechanics Laboratory.  We demonstrate the superiority of the nonlocal reconstruction formula for a large range of amplitudes. In addition, using the Implicit Function Theorem, we analyze the nonlocal formulation in order to demonstrate its solvability for the surface elevation given the pressure.

\section{A nonlocal formula relating pressure and surface elevation}

Consider Euler's equations describing the dynamics of the surface of an ideal fluid in two dimensions (with a one-dimensional surface):

\begin{align}\label{eqn:o1dEuler1}
\phi_{xx} + \phi_{zz} &=0,  & &(x,z) \in D,\\
\phi_z &=0, & &z = -h,  \label{eqn:o1dEuler2}\\
\eta_t + \eta_x\phi_x &=\phi_z,  &&z = \eta(x,t), \label{eqn:o1dEuler3}\\
\phi_t + \frac{1}{2} \left(\phi_x^2 + \phi_z^2\right) + g\eta &=0, && z = \eta(x,t),\label{eqn:o1dEuler4}
\end{align}

\begin{figure}[tb]
\hspace*{-0.2in}
\centering \includegraphics[width=.95\textwidth]{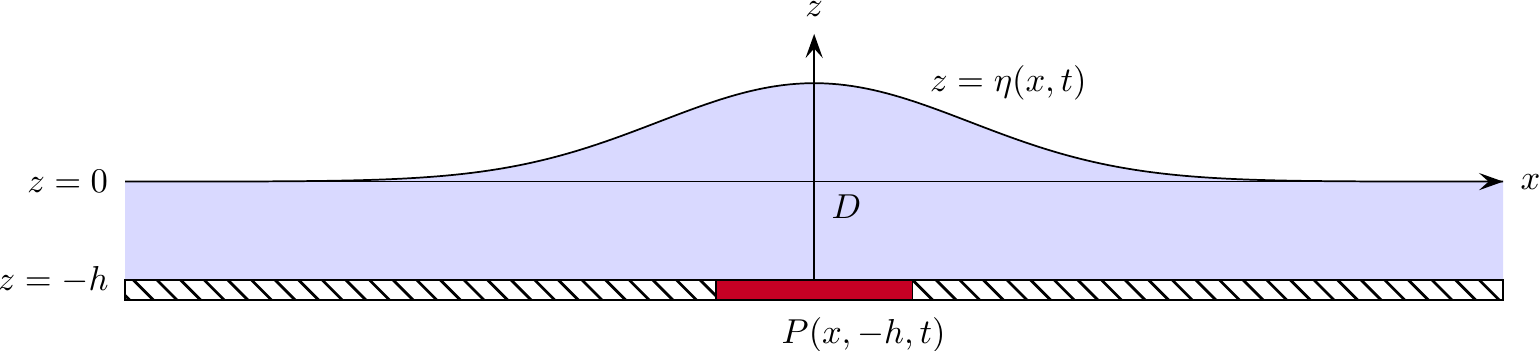}
{\caption{The fluid domain $D$ for the water wave problem. An idealized pressure sensor is indicated at the bottom. In our calculations the pressure measurement is assumed to be a point measurement.}\label{fig:fluidDomain}}
\end{figure}

\no where $\phi(x,z,t)$ represents the velocity potential of the fluid with surface elevation $\eta(x,t)$.
As posed, the equations require the solution of Laplace's equation inside the fluid domain $D$, see Figure~\ref{fig:fluidDomain}. If the problem is posed on the whole line $x\in \mathbb{R}$, we require that all quantities approach zero at infinity. If periodic boundary conditions are used then all quantities at the right end of the fluid domain are equal to those at the left end. Below we work with the whole line problem, stating only the results for the periodic case.

%Euler's equations as written above are challenging to work with directly: they are a free-boundary problem with %nonlinear boundary conditions specified at an unknown boundary. Many reformulations of the equations exist. We work %with that of Ablowitz, Fokas \& Musslimani \cite{AFM}. They introduced a nonlocal reformulation of the Euler %equations, valid for surface waves localized on the whole line or the whole plane. We repeat some of their %calculations, as they are essential for our final result.

Following \cite{zakharov}, let $q(x,t)$ represent the velocity potential at the surface $z = \eta(x,t)$, so that

\beq\la{qeq}
q(x,t) = \phi(x,\eta(x,t),t).
\eeq

\no Combining the above with equation (\ref{eqn:o1dEuler3}), we have

\begin{displaymath}
\phi_z = \eta_t + \left(q_x - \phi_z\eta_x\right)\eta_x,
\end{displaymath}

\no obtained by taking an $x$-derivative of \rf{qeq}. This allows us to solve directly for $\phi_z$ in terms of $\eta$ and $q$:

\beq\la{phiz}
\phi_z = \frac{\eta_t + \eta_xq_x}{1 + \eta_x^2}.
\eeq

\no Using (\ref{eqn:o1dEuler3}) again gives an expression for $\phi_x$, while taking a $t$-derivative of \rf{qeq} leads to an expression for $\phi_t$:

\beq
\phi_x = \frac{q_x - \eta_x\eta_t}{1 + \eta_x^2}, \quad \phi_t = q_t -  \frac{\eta_t\left(\eta_t + \eta_xq_x\right)}{1 + \eta_x^2} \label{eqn:VPotSurface}.
\eeq

\no Substituting these expressions into the dynamic boundary condition (\ref{eqn:o1dEuler4}) we find

\begin{equation}
q_t + \frac{1}{2}q_x^2 + g\eta - \frac{1}{2}\,\frac{(\eta_t + q_x\eta_x)^2}{1 +\eta_x^2} = 0,\label{eqn:surfaceBernoulli}
\end{equation}

\no after some simplification.
%The results above are all found in \cite{AFM}.

Next, we restrict to the case of a traveling wave moving with velocity $c$. We introduce $\xi = x - ct$, so that $x$ and $t$-derivatives become $\xi$-derivatives, the latter multiplied by $-c$. The Bernoulli equation \rf{eqn:surfaceBernoulli} becomes a quadratic equation in $q_\xi$:

\begin{equation}
-cq_\xi + \frac{1}{2}q_\xi^2 + g\eta - \frac{1}{2}\,\frac{\eta_\xi^2(q_\xi - c)^2}{1 +\eta_\xi^2} = 0.\label{eqn:surfaceBernoulliTWS}
\end{equation}

\no Solving this quadratic equation, we find

\begin{equation}\la{sweet}
q_\xi=c \pm\sqrt{(c^2-2g\eta)(1 + \eta_\xi^2)},
\end{equation}

\no where for $c>0$ we choose the $-$ sign, to ensure that the local horizontal velocity is less than the wave speed \cite{constantinstrauss}. Similarly, for $c<0$ the $+$ sign should be chosen.

Substituting this result into \rf{phiz} and \rf{eqn:VPotSurface}, we find

\begin{equation}\label{eqn:VPotSurfaceTWS}
\phi_\xi = c - \sqrt{\frac{c^2 - 2g\eta}{1 + \eta_\xi^2}}, \qquad \phi_z =  -\eta_\xi \sqrt{\frac{c^2 - 2g\eta}{1 + \eta_\xi^2}},
\end{equation}

\no where we have chosen $c>0$, without loss of generality. This simple calculations allows us to to express the gradient of the velocity potential at the surface directly in terms of the surface elevation.

Returning to the original coordinate system $(x,z,t)$, let  $Q(x,t) = \phi(x,-h,t)$, the velocity potential at the bottom of the fluid.  Inside the fluid, we know that the Bernoulli equation holds:

\beq\la{bernoulli}
\phi_t + \frac{1}{2}\left(\phi_x^2 + \phi_z^2\right) +gz +\frac{P(x,z,t)}{\rho} = 0, ~~~-h \leq z \leq \eta(x,t).
\eeq

\no Evaluating this equation at $z = -h$, we find

\begin{equation}
Q_t+\frac{1}{2}Q_x^2 -gh +\frac{P(x,-h,t)}{\rho} = 0.\label{eqn:BottomBernoulli}
\end{equation}

\no Moving to a traveling coordinate frame as before, we obtain a quadratic equation for $Q_\xi$. Solving for $Q_\xi$ we find

\begin{equation}\label{eqn:VPotBottomTWS}
Q_\xi  = c -\sqrt{c^2-2p},
\end{equation}

\no where $p(\xi)$ represents the non-static part of the pressure at the bottom in the traveling coordinate frame, scaled by the fluid density: $p(\xi) = P(x-ct, -h, t)/\rho - gh$. For consistency with our previous choice, we work with the $-$ sign again. Next, we connect the information at the surface with that at the bottom of the fluid.

Within the bulk of the fluid $D$,
\beq\la{travellaplace}
\phi_{\xi\xi}+\phi_{zz} = 0,
\eeq

\no where the boundary conditions given in (\ref{eqn:VPotSurfaceTWS}) and (\ref{eqn:VPotBottomTWS}) must also be satisfied. We can write the solution of this equation as

\beq\la{form}
\phi(\xi,z) = \frac{1}{2\pi}\displaystyle \int_{-\infty}^\infty e^{ik\xi}{\Psi}(k)\cosh\left(k\left(z +h\right)\right)\,dk,
\eeq

\no where the boundary condition for $\phi_z$ at $z =-h$ is satisfied. For the boundary condition at the bottom for $\phi_\xi$ we find

\beq
\frac{1}{2\pi}\int_{-\infty}^\infty ik e^{ik\xi} {\Psi}(k)\,dk=c-\sqrt{c^2-2p},
\eeq

\no so that

\begin{equation}
ik {\Psi}(k)=2\pi c \delta(k) - \mathcal{F}\left\lbrace\sqrt{c^2-2p}\right\rbrace(k),\label{eqn:psiHat}
\end{equation}
where $\delta(k)$ is the Dirac delta function and $\mathcal{F}$ denotes the Fourier transform: $\mathcal{F}\{y(\xi)\}(k)=\int_{-\infty}^\infty y(\xi)\exp(-ik\xi) d\xi$. Evaluating $\phi_\xi(\xi,z)$ at the surface $z = \eta$, we have

\begin{align*}
\phi_\xi(\xi,\eta) &= \displaystyle \frac{1}{2\pi} \int_{-\infty}^\infty e^{ik\xi}ik {\Psi}(k)\cosh\left(k\left(\eta +h\right)\right)\,dk\\
&=c-\frac{1}{2\pi}\int_{-\infty}^\infty e^{ik\xi} \cosh\left(k\left(\eta +h\right)\right)
\mathcal{F}\left\lbrace\sqrt{c^2-2p}\right\rbrace(k) \,dk.
\end{align*}

\no Using the boundary conditions given in (\ref{eqn:VPotSurfaceTWS}), we find the nonlocal relationship

\beq\la{awesome}
\sqrt{\frac{c^2 - 2g\eta}{1 + \eta_\xi^2}}=\frac{1}{2\pi}\int_{-\infty}^\infty e^{ik\xi} \cosh\left(k\left(\eta +h\right)\right)
\mathcal{F}\left\lbrace\sqrt{c^2-2p}\right\rbrace(k) \,dk.
\eeq

%\begin{eqnarray}
%&&\sqrt{\frac{c^2 - 2g\eta}{1 + \eta_\xi^2}} \label{eqn:SPRelationship}  =\\&&\int_{-\infty}^\infty %e^{ik\xi}\mathcal{F}\left(\sqrt{c^2 - 2gp(\xi)}\right)\cosh\left(k\left(\eta +h\right)\right)~dk.\nonumber
%\end{eqnarray}

Equation \rf{awesome} is the main result of this paper. It provides an implicit relationship between the surface elevation of a localized traveling wave $\eta(x)$ and the pressure measured at the bottom of the fluid $p(\xi)$. In the rest of this paper we investigate how this relationship may be used to compute $\eta(\xi)$ if $p(\xi)$ is known, and how different asymptotic formulas may be derived from it.

\newpage

{\bf Remarks.}

\begin{itemize}

\item In order to extend the above to periodic boundary conditions, we use the periodic generalization of the formulation of Ablowitz, Fokas, \& Musslimani {AFM}, see \cite{deconinckoliveras}. Following the steps outlined above, this allows for the derivation of a relation between the surface elevation of a periodic traveling wave and the pressure at the bottom:

    \beq\la{eqn:periodicawesome}
    \sqrt{\frac{c^2 - 2g\eta}{1 + \eta_\xi^2}}  = \frac{1}{2\pi}\sum_{k=-\infty}^\infty e^{ik\xi}
    \cosh\left(k\left(\eta +h\right)\right) \hat{P}_k,
    \eeq

    \no where $\hat{P}_k = \int_0^{2\pi} e^{-ik\xi}\sqrt{c^2-2p(\xi)}~d\xi$. In what follows, we will use either \rf{awesome} or \rf{eqn:periodicawesome}.

\item In the above, we have assumed there exists a smooth solution to the water wave problem (\ref{eqn:o1dEuler1}-\ref{eqn:o1dEuler4}).
    %Then $\phi_\xi$ is a harmonic function as well, which (assuming sufficient decay for $\phi$ in %\rf{travellaplace}) decays to zero at infinity.
    %{Add references to justify this or at least hint that it is possible, perhaps only for the travelling wave %case?}
    Given a speed $c$ and a non-hydrostatic pressure profile $p$ as inputs, we aim to solve \rf{awesome} for $\eta$. However, these inputs cannot be independent of each other: indeed arbitrary pressure profiles will not lead to surface elevations corresponding to solutions of (\ref{eqn:o1dEuler1}-\ref{eqn:o1dEuler4}).
    One expects that for a given speed $c$, there exists a unique surface elevation $\eta$ and associated pressure profile $p$. In order to back up this intuition, we require another relation between $\eta$, $p$ and $c$. Such a relation is found by taking a derivative with respect to $z$ of \rf{form} and equating the result to
    the right-hand side of the second equality in \rf{eqn:VPotSurfaceTWS}. Finally, \rf{eqn:psiHat} is used, resulting in
    %we impose the kinematic condition at the surface (\ref{KBC_NL}). Note that finding $\eta$ which satisfies %(\ref{NonlinearFormula}) does not guarantee the kinematic boundary condition.

\beq
\eta_x\sqrt{\frac{c^2-2g\eta}{1+\eta_x^2}} =\frac{-i}{2\pi} \int_{-\infty}^\infty e^{ikx}\sinh(k(\eta+h))\mathcal{F}\left\{c-\sqrt{c^2-2p}\right\}\!(k)dk.\label{NonlinearFormula2}
\eeq

\no The system of equations (\ref{awesome}) and (\ref{NonlinearFormula2}) may be solved to obtain both $\eta$ and $p$, given $c$. We will not pursue this issue further and content ourselves with establishing a map from $p$ to $\eta$.

For the purposes of the question considered in this paper, the above is not an issue: we assume that the given pressure originates from experimental observations and hence corresponds to the unique solution of (\ref{eqn:o1dEuler1}-\ref{eqn:o1dEuler4}), to the extent that the Euler equations provide an accurate model for the water wave problem.

\item {\bf Obtaining the pressure at the bottom from the surface elevation.}
An explicit nonlinear relationship for the pressure at the bottom in terms of the surface elevation may be obtained directly from the approach of Ablowitz, Fokas and Musslimani. Consider the relationship (the one-dimensional version of Equation $1.11$ in \cite{AFM})

\beq
\int_{-\oo}^\oo e^{ikx\pm |k|(\eta+h)}\left(\sgn(k)\eta_t - i\, q_x\right)\,dx = -i\int_{-\oo}^\oo e^{ikx}\phi_x(x,-h,t)dx, ~~~k\in \mathbb{R}_0.
\eeq

\no Changing to a moving frame of reference and substituting the expressions (\ref{sweet},\ref{eqn:VPotBottomTWS}) into this relation one obtains

\beq\la{exppressure}
\int_{-\oo}^\oo\!\! e^{ikx\pm |k|(\eta+h)}\left(\! -i c\,\sgn(k) \eta_x + \left(c - \sqrt{(c^2-2g\eta)(1+\eta_x^2)}\right)\right) dx =\! \int_{-\oo}^\oo\!\!\! e^{ikx}\left(c-\sqrt{c^2-2p}\right) dx.
\eeq

\no The right-hand side is essentially the Fourier transform of the quantity $c-\sqrt{c^2-2p}$. Inverting this transform, one may solve for the pressure at the bottom in terms of the surface elevation. Equation \rf{exppressure} provides an alternative to \rf{awesome} and can be used in its stead. The formula \rf{exppressure} is advantageous if one wishes to compute the pressure, given the surface elevation. The presence of a small-divisor problem (seen by linearizing the left-hand side integrand about $\eta=0$) when $c$ is near its shallow-water limit value $\sqrt{gh}$ indicates that \rf{awesome} is to be preferred over \rf{exppressure} to reconstruct the elevation $\eta$, given the bottom pressure $p$.

\item In \cite{constantinstrauss}, Constantin and Strauss derive various properties of the pressure underneath a traveling wave. It should be possible to re-derive these properties directly from \rf{awesome} or \rf{eqn:periodicawesome}. This is not pursued further here.

\end{itemize}

\section{Existence and uniqueness of solutions to the nonlinear formula}
	
In this section, we analyze \rf{awesome}. Among other results, using the Implicit Function Theorem, we show that the nonlocal relation \rf{awesome} gives rise to a well-defined map from the pressure profile to the surface elevation: given the pressure profile $p$ at the bottom, \rf{awesome} defines a unique surface elevation $\eta$. In other words, we can expect the asymptotic and numerical methods employed in the next sections to produce faithful approximations to the true solution.
%For a mathematical definition of ``reasonable'', please refer to Theorem~\ref{theo}.

Define the operator $F$, parameterized by $c\in\mathbb{R}$, by

\beq
F(\eta,p) =  c - \sqrt{\frac{c^2-2g\eta}{1+\eta_{x}^2}} - \frac{1}{2\pi}\int_{-\oo}^\oo e^{ikx}\cosh(k(\eta+h))\mathcal{F}\left\{c-\sqrt{c^2-2p}\right\}\!(k)dk.
\eeq

\no Note that $F(\eta,p)=0$ is equivalent to \rf{awesome}. Using the Implicit Function Theorem, we wish to show that the equation $F(\eta,p)=0$ has a solution profile $\eta$, given sufficiently small pressure $p$. We have that $F(0,0)=0$. In order to apply the Implicit Function Theorem, we need to define appropriate Banach spaces for which the operator $F$ is defined.
First we seek a suitable space for $\eta$. An obvious choice is $\eta\in C^1[\mathbb{R},\mathbb{R}]$, \emph{i.e.} $\eta$ is a continuously differentiable function which vanishes at infinity. This space is supplied with the usual norm:

\beq
\|\eta\|_{C^1} = \sup_{x\in\mathbb{R}}|\eta(x)| + \sup_{x\in\mathbb{R}}|\eta'(x)|.
\eeq

If $\|\eta\|_{C^1}<c^2/2g$ then $c - \sqrt{(c^2-2g\eta)/(1+\eta_{x}^2)}$ represents a continuous function of $x$. Hence we are motivated to define the image of $F$ in $C[\mathbb{R},\mathbb{R}]$. Consequently, we wish for

\[
\int_{-\oo}^\oo e^{ikx}\cosh(k(\eta+h))\mathcal{F}\left\{c-\sqrt{c^2-2p}\right\}\!(k)dk,
\]

\no to be a continuous function of $x$. For finite $\|\eta\|_{C^1}$, this nonlocal term is a continuous function of $x$ if

\beq
\int_{-\oo}^\oo \cosh(k(\|\eta\|_{C^1}+h))\left|\mathcal{F}\left\{c-\sqrt{c^2-2p}\right\}\!\!(k)\right|dk<\oo,\label{int_term_bound}
\eeq

\no and if the integrand of the nonlocal term is a continuous function of $x$ for every $k$ (see Theorem $2.27$ on pg.~56 of \cite{folland}). Let us consider the second condition, namely the continuity of the integrand. Since by assumption $\eta$ is continuous, the continuity in $x$ of the integrand requires

\beq
\sup_k \left|\mathcal{F}\left\{c-\sqrt{c^2-2p}\right\}\!\!(k)\right|<\oo.
\eeq

\no An application of the Cauchy-Schwarz inequality gives

\begin{align*}
\left|\mathcal{F}\left\{c-\sqrt{c^2-2p}\right\}\!(k)\right|\leq \left(\int_{-\oo}^\oo \frac{1}{1+|x|^2}dx\right)^{\frac1 2} \left(\int_{-\oo}^\oo (1+|x|^2)\left|c-\sqrt{c^2-2p}\right|^2 dx\right)^{\frac 1 2}.
\end{align*}

\no Hence, we impose the following condition on the pressure $p$:

\beq
\int_{-\oo}^\oo (1+|x|^2)\left|c-\sqrt{c^2-2p}\right|^2 dx<\oo.
\eeq

Next, we return to the first condition (\ref{int_term_bound}). Due to the presence of the hyperbolic cosine, we expect that it is necessary for $\mathcal{F}\{c-\sqrt{c^2-2p}\}\!(k)$ to have sufficient decay for large $|k|$. Let $M>h+\|\eta\|_{C^1}$. Starting from the integral in (\ref{int_term_bound}), we apply the Cauchy-Schwarz inequality again to find

\begin{align*}
&\int_{-\oo}^\oo \cosh(k(\|\eta\|_{C^1}+h))e^{-M|k|}e^{M|k|}\left|\mathcal{F}\left\{c-\sqrt{c^2-2p}\right\}\!\!(k)\right|dk\\ &\leq
 \left(\int_{-\oo}^\oo \left(\cosh(k(\|\eta\|_{C^1}+h))e^{-M|k|} \right)^2\!\!dk\right)^{\frac1 2}\left(  \int_{-\oo}^\oo e^{2M|k|}\left|\mathcal{F}\left\{c-\sqrt{c^2-2p}\right\}\!\!(k)\right|^2\!\!dk\! \right)^{\frac 1 2}\\&
\leq C\left( \int_{-\oo}^\oo e^{2M|k|}\left|\mathcal{F}\left\{c-\sqrt{c^2-2p}\right\}\!\!(k)\right|^2\!\!dk\! \right)^{\frac 1 2},
\end{align*}

\no for some constant $C$. Thus, if the conditions

\begin{align}
&&\int_{-\oo}^\oo e^{2M|k|}\left|\mathcal{F}\left\{c-\sqrt{c^2-2p}\right\}(k)\right|^2dk &<\oo\label{Condition1}\\
&\mbox{and}&\int_{-\oo}^\oo (1+|x|^2)\left|c-\sqrt{c^2-2p}\right|^2 dx&<\oo\label{Condition2}
\end{align}

\no hold, the function $F(\eta,p)(x)$ is continuous for $\eta\in C^1$.

Having found the conditions (\ref{Condition1}-\ref{Condition2}), we now determine an appropriate function space for $p$ so that they are satisfied. The following theorem due to Paley and Wiener (Theorem~4 pg.~7 of \cite{paleywiener}) is helpful.

\begin{thm}\label{PW1}
If $w(z)$ (with $z=x+iy$) is an analytic function in the strip $-\lambda\leq y\leq \mu$ where $\lambda, \mu>0$ and

\[
\int_{-\oo}^\oo |w(x+iy)|^2dx<\oo,\quad -\lambda\leq y\leq \mu,
\]

\no then there exists a measurable function $\hat{w}(k)$ such that

\[
\int_{-\oo}^\oo |\hat{w}(k)|^2e^{2\mu k}dk<\oo,\quad\quad \int_{-\oo}^\oo |\hat{w}(k)|^2e^{-2\lambda k}dk<\oo,
\]

\no and

\[
w(x+iy) = \lim_{A\to\oo} \int_{-A}^A\frac{1}{2\pi}\hat{w}(k)e^{ik(x+iy)}dk,\quad -\lambda\leq y\leq \mu
\]

\no where the limit is to be understood in the mean-square sense.
\end{thm}

In other words, the Fourier transform of $w(x)$ exists and it has decay as specified above. In particular, for any $M<\min\{\lambda,\mu\}$

\begin{align*}
\int_{-\oo}^\oo e^{2M|k|}|\hat{w}(k)|^2dk &= \int_{0}^\oo e^{2Mk}|\hat{w}(k)|^2dk + \int_{-\oo}^0 e^{-2Mk}|\hat{w}(k)|^2dk,\\
&=\int_{0}^\oo e^{2Mk}e^{-2\mu k}e^{2\mu k}|\hat{w}(k)|^2dk + \int_{-\oo}^0 e^{-2Mk}e^{2\lambda k}e^{-2\lambda k}|\hat{w}(k)|^2dk,\\
&\leq \int_{0}^\oo e^{2\mu k}|\hat{w}(k)|^2dk + \int_{-\oo}^0 e^{-2\lambda k}|\hat{w}(k)|^2dk<\oo.
\end{align*}

	The theorem implies that a sufficient condition for (\ref{Condition1}) to hold is that $c-\sqrt{c^2-2p}$ is an analytic function of $z$ within a strip of width at least $2M$ centered around the real axis and that it is square-integrable along lines parallel to the real axis within this strip. Of course, the presence of the square root is a hinderance to the analyticity of the function. Consequently, we require that $|p|<c^2/2$ everywhere in the strip. This condition also implies the square-integrability of the function if $p$ is square-integrable. Indeed, the function

\[
f(z) = c - \sqrt{c^2-2z},
\]

\no is analytic (and hence Lipschitz) in a neighborhood of the origin for which $|z|<c^2/2$. Hence, for all $z_1$, $z_2$ in such a neighborhood of the origin we have

\beq
|f(z_1)-f(z_2)|\leq C |z_1-z_2|,
\eeq

\no for some constant $C$. In particular, since $f(0)=0$,

\beq
|f(z)|\leq C|z|,
\eeq

\no uniformly for all $|z|\leq \delta<c^2/2$, \emph{i.e.}, the constant $C$ is independent of $z$. Next, consider the function $f(p(z))$, where $p(z)$ is a function which is analytic and bounded in the strip of width $2M$. This implies

\beq
|f(p(z))|\leq C|p(z)|.
\eeq

\no As above, the constant $C$ is independent of $p(z)$ and thus of $z$, provided $|p(z)|<c^2/2$. Thus the square-integrability of $p(z)$ implies the square-integrability of $c-\sqrt{c^2-2p}$ for $|p|\leq\delta<c^2/2$ for every $z$ in the strip. Thus, if $p(z)$ is an analytic function in the strip of width at least $2M$, square-integrable along lines parallel to the real axis, and bounded in the strip, the first condition (\ref{Condition1}) holds.

Another theorem due to Paley and Wiener (Theorem 2 pg 5 of \cite{paleywiener}) allows us to bound $|p|$ in terms of the $L^2$-norm.

\begin{thm}\label{PW2}
Let $w(z)$ be analytic in the strip $-\lambda\leq y\leq \mu$ with $\mu,\lambda>0$ and

\beq
\int_{-\oo}^\oo |w(x+iy)|^2dx<\oo,\quad -\lambda\leq y\leq \mu,
\eeq

\no then for any $z$ in the interior of the region

\beq
w(z) = \frac{1}{2\pi}\int_{-\oo}^\oo \frac{w(x+i\mu)}{x+i\mu-z}dx - \frac{1}{2\pi}\int_{-\oo}^\oo \frac{w(x-i\lambda)}{x-i\lambda-z}dx.
\eeq

\no In particular, an application of the Cauchy-Schwarz inequality shows that for any $y\in[-\lambda+\epsilon,\mu-\epsilon]$ with $\epsilon>0$, $w(z)$ is bounded in terms of the $L^2$ norms of $w(x+i\mu)$ and $w(x-i\lambda)$.
\end{thm}

	Collecting these ideas, we choose the pressure $p$ to be in the space of analytic functions in the symmetric strip of width $2M$ about the real axis such that

\beq\la{space}
\int_{-\oo}^\oo (1+|x|^2)|p(x+iy)|^2dx<\oo,\quad -M\leq y\leq M.
\eeq

\no Note that this condition guarantees that the second condition \rf{Condition2} is also satisfied.
Let $H_M$ denote the space defined by \rf{space}. It is endowed with the norm

\beq
\|p\|_{H_M}=\sup_{|y|\leq M}\left[\int_{-\oo}^\oo (1+|x|^2)|p(x+iy)|^2dx\right]^{1/2}.\label{HM_norm}
\eeq

\no We claim that $H_M$ is a Banach space. Indeed, with the obvious definitions of addition and scalar multiplication for elements $p\in H_M$, $H_M$ is a vector space. It is straightforward to verify that (\ref{HM_norm}) defines a norm. Thus, it remains to verify completeness. Let $\lbrace f_k \rbrace$ be a Cauchy sequence in $H_M$. With the norm above, this sequence converges to a complex-valued function $f$ defined on the strip since for fixed $y$, $\lbrace f_k \rbrace$ defines a Cauchy sequence in the space $L^2$ with weight $(1+|x|^2)$ and has the limit $f(\cdot,y)$ in this space for each $y$. We define the function

\beq
f(x,y) = \lim_{k\to\oo} f_k(x+i y),\mbox{ y fixed}.
\eeq

\no Since $\lbrace f_k \rbrace$ is a Cauchy sequence, for every $\epsilon>0$ there exists an $N$ such that for $n,k\geq N$
\beq
\|f_n - f_k\|_{H_M}\leq \epsilon.
\eeq

\no This implies

\beq
\int_{-\oo}^\oo (1+|x|^2)|f_n(x+iy)-f_k(x+iy)|^2dx\leq \epsilon,
\eeq

\no for every $|y|\leq M$. Letting $k\to\oo$ in the above integral we obtain

\beq
\int_{-\oo}^\oo (1+|x|^2)|f_n(x+iy)-f(x,y)|^2dx\leq \epsilon,
\eeq

\no for every $|y|\leq M$ and thus $f_n\to f$ in the $H_M$ norm. Theorem \ref{PW2} implies the pointwise bound

\beq
|w(x+iy)|\leq C \|w(x+iy)\|_{H_M},
\eeq

\no for $x+iy$ in the interior of the strip. Consequently, convergence in $H_M$ implies uniform convergence on compact subsets of the strip. From Morera's Theorem, it follows that the Cauchy sequence of analytic functions $f_k$ converges to an analytic function, thus the space $H_M$ is complete.

We now state the theorem for the existence of a map from the pressure beneath a traveling wave to the surface elevation of the wave.

\begin{thm}\la{theo}
Let $p$ and $\eta$ be the bottom pressure and surface elevation, respectively, obtained by solving the Euler equations augmented with (14). Assume that $p \in H_{M+\epsilon}$ for some $M>h$, $\epsilon>0$ and that $\|\eta\|_{C^1}<\min[M-h,c^2/2g]$. Then for fixed $c\neq 0$ and sufficiently small $p$, the equation

\[
c - \sqrt{\frac{c^2-2g\eta}{1+\eta_{x}^2}} = \frac{1}{2\pi}\int_{-\oo}^\oo e^{ikx}\cosh(k(\eta+h))\mathcal{F}\left\{c-\sqrt{c^2-2p}\right\}\!\!(k)dk,
\]

\noindent has a solution $\eta$.
%This solution may be written as $\eta=\nu(p)$, where $\nu: p\rightarrow \eta$ is a continuously differentiable map.
Further, if $p$ is the true pressure consistent with (\ref{eqn:o1dEuler1}-\ref{eqn:o1dEuler4},\ref{bernoulli}), then the only solution for $\eta$ is that which solves the stationary water wave problem with speed $c$.

%holds for $\eta=\nu(p)$, for $p$ sufficiently small and fixed $c\neq 0,h>0$. Here $\nu:p\to \eta$ is a continuously %differentiable map. Further, the only values of $\eta$ which satisfy the above equation for the true pressure $p$ are %those obtained from the Euler equations and are of the form $\nu(p)$.
\end{thm}

{\bf Proof.}
Let $M>h$ and $\epsilon>0$. By Theorem~\ref{PW2} there is a ball $V$ around the origin in $H_{M+\epsilon}$, \emph{i.e.} $V=\{p\in H_{M+\epsilon}: \|p\|_{H_{M+\epsilon}}<\delta\}$, such that

\[
\sup_{|y|\leq M+\epsilon/2}|p(x+iy)|\leq S<\frac{c^2}{2}.
\]

\noindent By definition, $p\in V$ is bounded, analytic and square integrable along lines parallel to the real axis. Then the function $c-\sqrt{c^2-2p}$ is also analytic and square integrable along lines parallel to the real axis in a strip of width $2(M+\epsilon/2)$ symmetric with respect to the real axis. Using Theorem \ref{PW1},

\[
\int_{-\oo}^\oo e^{2M|k|}\left|\mathcal{F}\left\{c-\sqrt{c^2-2p}\right\}\!\!(k)\right|^2dk<\oo.
\]

\noindent For $p\in V$, following the discussion preceeding Theorem~\ref{PW1}, define the reconstructed functions $\phi^{R,x}$ and $\phi^{R,z}$ as

\begin{align}\label{phirx}
\phi^{R,x}(x,z;p) &= \frac{1}{2\pi}\int_{-\oo}^\oo e^{ikx} \cosh(k(z+h))\mathcal{F}\left\{ c-\sqrt{c^2-2p}\right\}(k) dk,\\\label{phirz}
\phi^{R,z}(x,z;p) &= \frac{1}{2\pi}\int_{-\oo}^\oo -ie^{ikx} \sinh(k(z+h))\mathcal{F}\left\{ c-\sqrt{c^2-2p}\right\}(k) dk.
\end{align}

\noindent Then  $\phi^{R,x}$ and $\phi^{R,z}$ are harmonic (and thus smooth) for all $x\in \mathbb{R}$ and $|z+h|<M$.
Let $R=\min\left(M-h,c^2/2g\right)$ and define the ball $U=\lbrace \eta\in C^1:\|\eta\|_{C^1}<R\rbrace$. Then $G:U\times V \to C^1[\mathbb{R},\mathbb{R}]$, defined as

\[
G(\eta,p)=-\phi^{R,x}(x,\eta;p) + \frac{1}{2c} [\phi^{R,x}(x,\eta;p)]^2 + \frac{1}{2c} [\phi^{R,z}(x,\eta;p)]^2 + \frac{g}{c}\eta,
\]

\noindent is a continuously differentiable function with $G(0,0)=0$, as is readily verified by computing its second variation evaluated at $(\eta, p)=(0,0)$. The Fr\'{e}chet derivative of $G$ with respect to $\eta$ at the origin is

\[
G_\eta(0,0)v=\frac{g}{c}v,\quad\quad v\in C^1.
\]

\noindent The Fr\'{e}chet derivative $G_\eta(0,0)$ is an isomorphism on $C^1$. Hence the Implicit Function Theorem applies and there exists a continuously differentiable map $\nu : p \to \eta$ such that $G(\nu(p),p)=0$ for all sufficiently small $p$.

Next, we show that if $p$ is the pressure consistent with the traveling water wave problem with velocity $c$, then $\nu(p)$ is indeed the corresponding water wave surface elevation. This is achieved by establishing that the reconstructed functions $\phi^{R,x}$ and $\phi^{R,z}$ are the horizontal and vertical fluid velocities $\phi_x$ and $\phi_z$, respectively.

Let $D=\lbrace -\oo<x<\oo, -h<z<\eta\rbrace$, as before, where $\eta$ represents the solution for the surface elevation of the traveling water wave problem with velocity potential $\phi$. Hence $\phi$ is harmonic in $D$. It is possible to harmonically extend $\phi$ to $\bar D=\lbrace-\oo<x<\oo,-\eta-2h<z<\eta\rbrace$ by reflecting the problem across the mirror line $z=-h$. Thus $\phi_x$ is harmonic in $\bar D$.

If $p$ is the pressure corresponding to the solution $(\phi, \eta)$ of the water wave problem through (\ref{bernoulli}), then at $z=-h$, $\phi^{R,x}$ and its normal derivative take the same values as $\phi_x$ and its normal derivative, respectively. The Cauchy-Kowalevski Theorem for Laplace's equation \cite{evans} implies that $\phi_x=\phi^{R,x}$ in a region near $z=-h$. But then $\phi_x$ and $\phi^{R,x}$ and all their derivatives are equal in this region. This implies that $\phi_x=\phi^{R,x}$ in $D$, by analytic continuation. A similar argument shows that $\phi_z=\phi^{R,z}$ in $D$:  to determine $\phi_{zz}$ at $z=-h$ we use the fact that due to the extension to $\bar D$, $\phi$ is harmonic on $z=-h$. In addition, from (\ref{phirx}) and (\ref{phirz}), $\phi^{R,x}$ and $\phi^{R,z}$ harmonically extend up to $z=M-h>\|\eta\|_{\oo}$. Since $\phi_x=\phi^{R,x}$ and $\phi_z=\phi^{R,z}$ in $D$, we can harmonically extend $\phi_x$ and $\phi_z$ up to $z=M-h$. This implies that $\phi^{R,x}=\phi_x$ and $\phi^{R,z}=\phi_z$ at $z=\eta$. Hence, from (\ref{eqn:o1dEuler4}) in the traveling frame of reference,

\begin{align*}
-c\phi_x + \frac 1 2 \phi_x^2 + \frac 1 2 \phi_z^2 + g\eta &= 0,\quad z=\eta\\
\Rightarrow -c\phi^{R,x} + \frac 1 2 (\phi^{R,x})^2 + \frac 1 2 (\phi^{R,z})^2 + g\eta &= 0,\quad z=\eta,\\
\Rightarrow G(\eta,p) &=0.
\end{align*}

\noindent From the Implicit Function Theorem, for small $p$ all solutions $\eta$ to $G(\eta,p)=0$ are of the form $\eta=\nu(p)$, where $\nu: p\rightarrow \eta$ is a $C^1$ map. Hence

\[
\phi^{R,x}(x,\nu(p);p)=\phi^{R,x}(x,\eta;p)=\phi_x(x,\eta)=c - \sqrt{\frac{c^2-2g\eta}{1+\eta_x^2}} = c - \sqrt{\frac{c^2-2g\nu(p)}{1+\nu^2_x(p)}},
\]

\noindent where we used \rf{eqn:VPotSurfaceTWS}. In other words, there are functions $\eta$ which depend continuously on the true pressure $p$ such that (\ref{awesome}) is true.

Next, assume there exists a different solution $\tilde{\eta}\in U \subset C^1, \tilde{\eta}\neq \eta$ such that

\[
c - \sqrt{\frac{c^2-2g\tilde{\eta}}{1+\tilde{\eta}_x^2}} = \phi^{R,x}(x,\tilde{\eta};p),
\]

\noindent for all $x$,
where the pressure $p$ is the pressure corresponding to the traveling water wave problem with velocity $c$. As before, $\phi^{R,x}(x,\tilde{\eta};p) = \phi_x(x,\tilde{\eta})$ and thus

\[
c - \sqrt{\frac{c^2-2g\tilde{\eta}}{1+\tilde{\eta}_x^2}} = \phi_x(x,\tilde{\eta}),
\]

\noindent for all $x$. However, for a fixed $c$, the water wave problem has a unique traveling wave solution \cite{amick}, which is contradicted by the statement that $\eta\neq \tilde \eta$.  Thus the only solutions $\eta$ of (\ref{awesome}) associated with the pressure $p$ are the traveling wave solutions of the Euler equations.

\section{Asymptotic approximations} \label{sec:asymptotics}

In this section we derive a variety of asymptotic approximations to the pressure as a function of the surface elevation. Given the complexity of \rf{awesome}, such approximations are especially useful. In the sections below, we compare the results for the pressure obtained using \rf{awesome}, with those obtained from \rf{archie} and \rf{engr}, as well as some asymptotic formulas obtained here.

We introduce the nondimensional quantities $\xi^*$, $z^*$, $\eta^*$ and $k^*$:

\beq
\xi^* = \xi/L, \quad z^* = z/h, \quad \eta^* = \eta/a, \quad k^* = L k, \quad c^*=c/\sqrt{gh}, \quad p^*=p/gh,
\eeq

\no where $L$ is a typical horizontal length scale, and $a$ is the amplitude of the surface wave. From  (\ref{eqn:o1dEuler1}--\ref{eqn:o1dEuler4}), a nondimensional version of  (\ref{awesome}) is found to be

\begin{equation}\label{eqn:awesomeNonDim}
\sqrt{\frac{c^2 - 2\epsilon\eta}{1 + (\epsilon\mu\eta_\xi)^2}}  = \frac{1}{2\pi}
\int_{-\infty}^\infty e^{ik\xi}\mathcal{F}\left\{\sqrt{c^2 - 2\epsilon p(\xi)}\right\}(k)\cosh\left(\mu k\left(1 +\epsilon \eta \right)\right)\,dk,
\end{equation}

\no where $\epsilon = a/h$ and $\mu = h/L$. The $^*$'s have been omitted to simplify the notation. This form of the nonlocal relation is our starting point to derive various approximate results. The two parameters $\epsilon$ and $\mu$ provide many options for different asymptotic expansions: we may assume small amplitude waves ($\epsilon \ll \mu$), or we may assume a long-wave approximation ($\mu \ll \epsilon$), or we may balance both effects as in a Korteweg-de Vries (KdV)-type approximation (see \cite{as}, for instance).

\subsection{The small-amplitude approximation: SAO1 and SAO2}

If we expand $\eta$ in powers of $\epsilon\ll 1$, assuming that $\epsilon \ll \mu$, we recover at leading order the approximation

\begin{equation}\label{eqn:Small_Amp_Order_1}
\displaystyle \eta(\xi) = \mathcal{F}^{-1}\big\lbrace \cosh(\mu k)\widehat p (k) \rbrace\big\rbrace + \mathcal{O}\left(\epsilon\right),
\end{equation}

\no where $\widehat p (k)=\mathcal{F}\{p\}(k)$. Ignoring the $\mathcal{O}\left(\epsilon\right)$ term, \rf{eqn:Small_Amp_Order_1} is the nondimensional version of \rf{engr}.
This demonstrates that \rf{awesome} is consistent with the frequently used \rf{engr}, and we are able to recover such formulas in a consistent manner using the single equation \rf{awesome}, instead of having to work with the full set of equations of motion.  For the remainder of this paper, we will refer to the model \rf{eqn:Small_Amp_Order_1} (solved for the pressure) as SAO1 (Small-Amplitude, Order 1).

If we proceed to higher order in $\epsilon$ we find the presumably more accurate approximation

\begin{equation}\label{eqn:Small_Amp_Order_2}
\eta(\xi) = \eta_0(\xi) + \epsilon \eta_1(\xi) + \mathcal{O}(\epsilon^2),
\end{equation}

\no where

\begin{align}
\eta_0(\xi) =& \mathcal{F}^{-1}\big\lbrace \cosh(\mu k)\widehat{p}(k)\rbrace\big\rbrace,\\
\eta_1(\xi) =& -\frac{c^2\mu^2}{2}{\eta_{0}}_\xi^2- \frac{1}{2c^2}\eta_0^2 + \mu\eta_0\mathcal{F}^{-1}\lbrace k\widehat{p}(k)\sinh(\mu k)\rbrace  + \frac{1}{2c^2}\mathcal{F}^{-1}\lbrace\widehat{p^2}(k)\cosh(\mu k)\rbrace.
\end{align}

The formula \rf{eqn:Small_Amp_Order_2} provides a new, explicit, higher-order approximation for the surface elevation $\eta(\xi)$ in terms of the pressure $p(\xi)$ and the traveling wave speed $c$, assuming a small-amplitude approximation.  For the remainder of this paper, we will refer to this model as SAO2.

\subsection{The KdV Approximation: SWO1 and SWO3}

Alternatively, we can balance the parameters $\mu$ and $\epsilon$ so that $\mu = \sqrt{\epsilon}$. This is the KdV approximation, see \cite{AFM, as}).  At leading order, we recover the simplest approximation that the surface elevation equals the pressure:

\begin{equation}\label{eqn:KdV_Order_1}
\eta(\xi) = p(\xi) + \mathcal{O}(\epsilon).
\end{equation}

This equation is exactly the hydrostatic approximation \rf{archie} in dimensionless variables; we will refer to this model as SWO1 (Shallow Water, Order 1). Continuing the aproximations to higher order (up to order $\epsilon^3$), we find

\begin{equation}\label{eqn:KdV_Order_3}
\eta(\xi) = p - \frac{\epsilon}{2}\ppn{2}{p}{\xi} + \epsilon^2\left( \frac{1}{24}\ppn{4}{p}{\xi} \!-\! p\ppn{2}{p}{\xi} \!-\! \frac{1}{2}\left(\pp{p}{\xi}\right)^2\left(c^2+\frac{1}{c^2}\right) \right) + \mathcal{O}(\epsilon^3).
\end{equation}

\no We refer to this approximation as SWO3 (Shallow Water, Order 3).

\vspace*{0.1in}

{\bf Remarks.}

\begin{itemize}

\item For $\eta \in C^1$ and $p\in H_{M+\delta}$, it is possible to prove the analyticity of \rf{eqn:awesomeNonDim} in $\epsilon$ and $\mu$. Here, as before, $M$ is related to the size of a symmetric strip around the real $\xi$ axis. This analyticity serves to validate the asymptotic approximations derived above, as being obtained through a process that gives the first few terms of a convergent series.

\item It appears to be a big restriction that the nonlocal formula \rf{awesome} and the approximations derived above require a traveling wave profile. As shown in the next section, good results are also obtained for waves that are not merely traveling at constant speed. For waves in shallow water, excellent agreement is often obtained by using $c=1$ (or $c=\sqrt{gh}$, returning to the dimensional version), which may be regarded as the zero-order approximation of an asymptotic series for $c$ in terms of $\epsilon$.

\item Using the procedures outlined in this section, the reader will find it straightforward to derive yet different approximations for the surface elevation in terms of the pressure measured at the bottom. For instance, one may consider a shallow-water approximation without imposing that the waves are of small amplitude, {\em i.e.}, $\mu\ll\epsilon<1$, {\em etc}.

\end{itemize}

\subsection{A heuristic formula: SAO2h}

The transfer function approach \rf{engr} is very successful for a variety of reasons: (i) it is quite accurate, as is illustrated in the next few sections. This statement remains true to a varying degree for waves of relatively high amplitude; (ii) the most complicated aspect of using the formula is the computation of two Fourier transforms; and (iii) the formula applies to waves that are not necessarily traveling with constant speed. This is a consequence of the linearization that led to \rf{engr}: each individual linear wave is traveling at constant speed, but typically their superposition is not.

In this section we derive a different formula for the reconstruction of the surface elevation from the pressure at the bottom. This formula is obtained somewhat heuristically, and its justification rests on the fact that it agrees extremely well with both numerical and experimental data. Furthermore, its use requires the computation of only three Fourier transforms, and the velocity $c$ does not appear in the final result. As a consequence, even though the derivation does not justify this, it is straightforward to apply to non-traveling wave profiles, where it performs very well. As for the other formulas above, the numerical and experimental results are presented below.

An equivalent form of the nondimensional nonlocal equation \rf{eqn:awesomeNonDim} is

\begin{equation}\label{eqn:awesomeNonDim2}
1-\sqrt{\frac{1 - 2\epsilon\eta/c^2}{1 + (\epsilon\mu\eta_\xi/c)^2}}  = \frac{1}{2\pi}
\int_{-\infty}^\infty e^{ik\xi}\hat P(k, \epsilon)\cosh\left(\mu k\left(1 +\epsilon \eta \right)\right)\,dk,
\end{equation}

\no where

\beq
\hat P(k,\epsilon)=\mathcal{F}\left\{1-\sqrt{1 - 2\epsilon p(\xi)/c^2}\right\}(k).
\eeq

\no So as to consider a small-amplitude approximation, we expand this equation in powers of $\epsilon$. However, we do not expand $\hat P(k,\epsilon)$ at this point. Proceeding this way and retaining only first-order terms in $\epsilon \eta$ and $\epsilon \eta_\xi$, we find

\begin{align}\nonumber
&&\frac{\epsilon \eta}{c^2}&=\frac{1}{2\pi}
\int_{-\infty}^\infty e^{ik\xi}\hat P(k, \epsilon)\left(
\cosh(\mu k)+\epsilon \mu \eta k \sinh(\mu k)
\right)\,dk\\\la{sweat}
&\Rightarrow&\epsilon \eta&=\ds \frac{\frac{1}{2\pi}
\int_{-\infty}^\infty e^{ik\xi}\hat P(k, \epsilon)\cosh(\mu k)\,dk}{\frac{1}{c^2}-\frac{\mu}{2\pi}
\int_{-\infty}^\infty e^{ik\xi}\hat P(k, \epsilon) k \sinh(\mu k)\,dk}.
\end{align}

Next, we expand $\hat P(k,\epsilon)$ in $\epsilon$, omitting terms of order $\epsilon^2$ and higher. We obtain

\beq\la{inter}
\hat P(k,\epsilon)=\frac{\epsilon}{c^2}\hat p.
\eeq

\no Substitution of \rf{inter} in \rf{sweat} results in

\beq\la{secret}
\eta = \frac{\mathcal{F}^{-1}\left\{\hat p(k)\cosh(\mu k)\right\}}{1-\epsilon \mu \mathcal{F}^{-1}\left\{\hat p(k)\,k \sinh(\mu k)\right\}}.
\eeq

\no As stated above, this reconstruction formula does not depend on $c$, and its application requires the computation of a mere three Fourier transforms. This can be contrasted, for instance, with the formula SAO2 which also uses a small-amplitude approximation. That formula requires the computation of five Fourier transforms and has explicit dependence on $c$. In fact, if one were to expand \rf{secret} in powers of $\epsilon$ one would find at order $\epsilon^0$ the transfer function formula \rf{engr}, and at order~$\epsilon^1$ the result SAO2 with all $c$-dependent terms omitted. We refer to the results obtained using \rf{secret} as SAO2h.

\section{Comparisons of the Different Approaches}

In this section, we present numerical results for the reconstruction of the surface elevation using the various relationships derived in Section~\ref{sec:asymptotics} for both numerical and experimental pressure data.  For the comparison using numerical data, we use previously computed periodic traveling waves solutions from \cite{deconinckoliveras}.  By using the exact pressure underneath the traveling wave, we attempt to reconstruct the surface elevation.  The same is done for various sets of experimental data obtained from the one-dimensional wave tank at the William Pritchard Fluids Laboratory at Penn State University.

\subsection{Comparison of the Different Approaches Using Numerical Data}

Using traveling wave solutions with periodic boundary conditions as calculated in \cite{deconinckoliveras}, we determine the pressure at the bottom using \rf{awesome} as follows. Without loss of generality, we assume that the solutions are $2\pi$ periodic.  For a given traveling wave solution profile specified by $(\eta_{true}(\xi),c_{true})$, the pressure $p(\xi)$ is obtained by equating the $k$-th Fourier coefficient of both the right- and left-hand side of (\ref{awesome}) for $k = -N\ldots N$, using a sufficiently high value of $N$.  This results in a linear system of algebraic equations for the coefficients of the Fourier series of $\sqrt{c^2 - 2p}$.  Using this truncated Fourier series, we may solve directly for $p(\xi)$ in terms of the given solution set  $(\eta_{true}(\xi),c_{true})$. Note that \rf{exppressure} offers a numerically equivalent alternative for computing $p(\xi)$.

Our goal is to reconstruct the surface elevation from the thus computed pressure at the bottom, using the various formulas given above. The asymptotic formulas given in the previous section do not require anything more complicated than a fast Fourier transform. The solution of the nonlocal equation \rf{awesome} is obtained using a pseudo-spectral method with differentiation carried out in Fourier space, while multiplication is carried out in physical space. We reconstruct $\eta$ by using a nonlinear solver such as a Gauss-Newton or Dogleg method \cite{dennis, nocedalwright} with an error tolerance of $10^{-14}$.  As an initial guess for our nonlinear solver, we use the approximation from (\ref{eqn:Small_Amp_Order_2}). Of course, the result obtained from the nonlocal equation should return the original surface elevation profile used to generate the pressure data, within machine precision. This provides a validation for the various numerical methods used. Here we compare the results from the asymptotic formulas of the previous section and evaluate their different errors.

\begin{figure}[tb]
\begin{center}
\begin{tabular}{c}
\includegraphics[width=.95\textwidth]{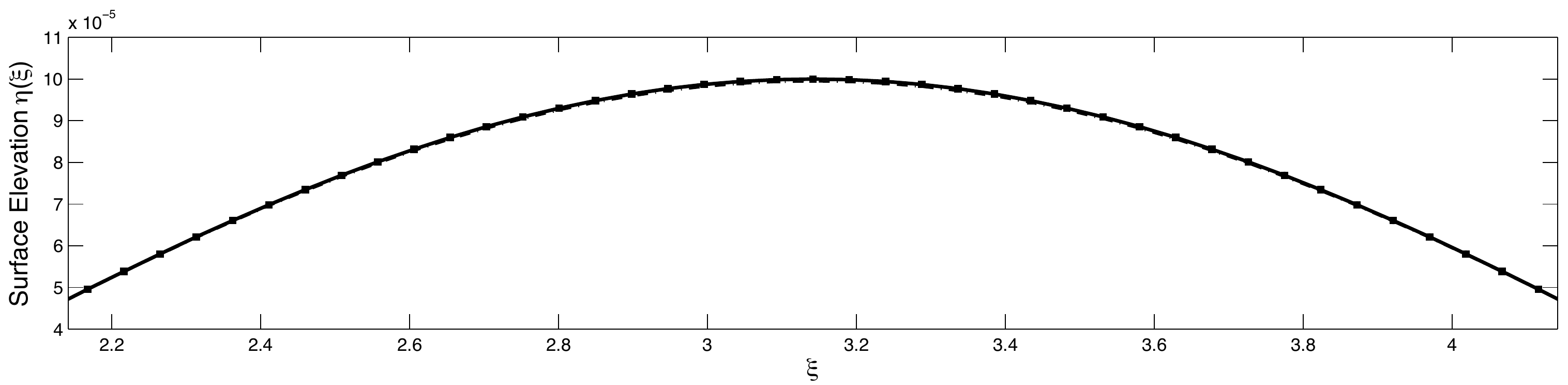}\\
(a)\\
\includegraphics[width=.95\textwidth]{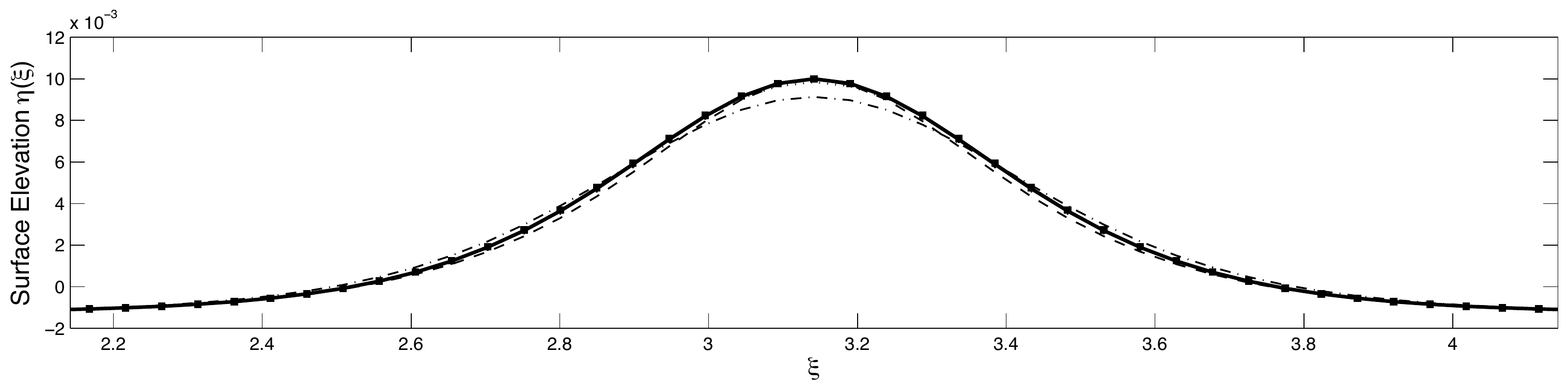}\\
(b)
\end{tabular}
\end{center}
{\caption{Reconstruction of the surface elevation from pressure data based on numerical experiments for $h = 0.1$, $g = 1$, $\rho=1$ and $L = 2\pi$.  Amplitudes are $a = 0.0001$ (a) and $a = 0.0056$. (b). No legend is included: all approximations including the nonlocal formula \rf{awesome} result in indistinguishable curves, except for the hydrostatic approximation SWO1, which displays a significant discrepancy for the bottom numerical experiment.}\label{fig:NumericReconstruction}}
\end{figure}

Using the parameter values $h = .1$, $g = 1$, $\rho=1$ and $L = 2\pi$, we reconstruct the solution for various solution amplitudes and speeds.  For solutions of small amplitude (say $ak = .0001$), we see that the reconstructions using all methods are in excellent agreement with the true surface wave elevation, see Figure~\ref{fig:NumericReconstruction}a. However, even for waves with amplitudes less than 15\% of the limiting wave height as given by \cite{cokelet} it becomes clear that certain approximations yield better results than others, see Figure~\ref{fig:NumericReconstruction}b.  In particular, while the nonlocal formula and the higher-order asymptotic formula reconstruct the wave profile well, the hydrostatic approximation SWO1 reconstruction fails to reproduce an accurate reconstruction of the peak wave height.

\begin{figure}[tb]
\centering \includegraphics[width=.95\textwidth]{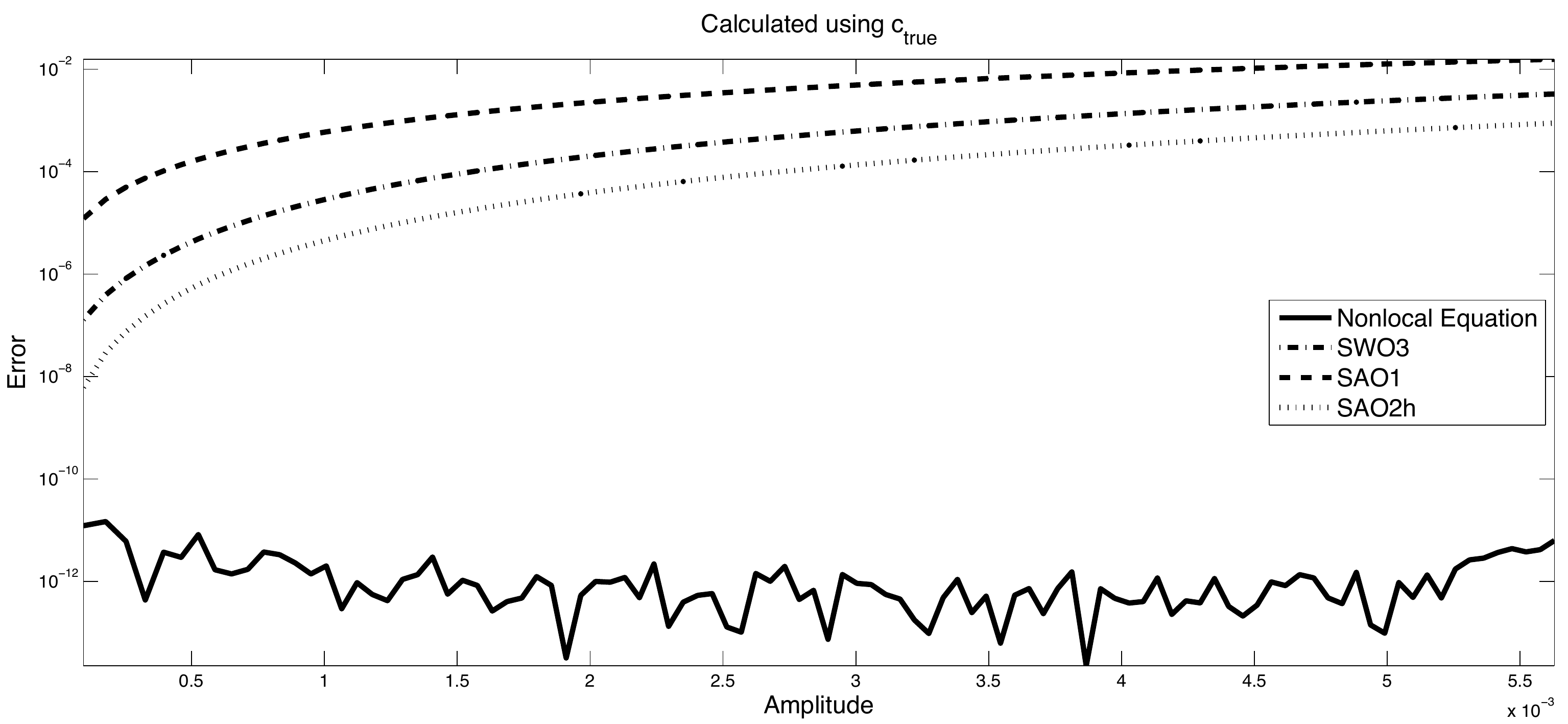}
{\caption{Plot of the relative error \rf{relerror} in the reconstructed surface elevation $\eta_r$ as a function of the amplitude of $\eta_{true}$ using the true value of the wave speed $c$. The asymptotic approximation SAO2 is not included in this figure. It is more costly to compute than its heuristic counterpart SAO2h, which yields better results.}\label{fig:NumericErrorRealC}}
\end{figure}

To demonstrate how the error changes as a function of the wave amplitude (or nonlinearity), we compute the relative error

\beq\la{relerror}
\textrm{error} = \frac{||\eta_{true} - \eta_r||_\infty}{||\eta_{true}||_\infty},
\eeq

\no where $\eta_{true}$ represents the expected solution and $\eta_r$ represents the reconstructed solution.  For the same nondimensional parameters as before, we calculate the error as a function of increasing peak wave height demonstrated in Figure~\ref{fig:NumericErrorRealC}. As seen there, the error in all approximations grows as the amplitude of the Stokes wave increases. Figure~\ref{fig:NumericErrorRealC} includes only solutions of small amplitude. If solutions of larger amplitude are considered, the discrepancies between the different approximations grow, as shown in Table~\ref{moderate}.

This table illustrates the large error generated by the lower-order methods SAO1 and SW01 for waves which are no more than 55\% of the limiting wave height as calculated in \cite{cokelet}. Even for waves which are 50\% of the limiting wave height, the relative error \rf{relerror} of the commonly used transfer function reconstruction SAO1 exceeds 15\%. In contrast, the higher-order methods SA02, SWO3, and SA02h consistently yield more accurate results.
It is also clear from the table that even for large amplitude waves, the nonlocal formula \rf{awesome} (or more precisely for these numerical data sets, its periodic analogue \rf{eqn:periodicawesome}) provides a practical means to reconstruct the surface elevation from pressure data measured along the bottom of a fluid, at least in this numerical data setting. Below we establish the same using physical experiments.

%This table considers moderate-amplitude waves with surface elevations which are no more than 55\% of the limiting wave %height as calculated in \cite{cokelet}. Table~\ref{moderate} illustrates that some approximations show errors %exceeding 15\% for such waves.
%As is obvious from both Figure~\ref{fig:NumericErrorRealC} and Table~\ref{moderate}, the numerical reconstruction %using the nonlocal equation barely suffers from this amplitude dependence. In fact, the only reason for the increase %in the error in the last column of Table~\ref{moderate} is due to the fact that the same number of Fourier modes was %used for the computation of all traveling wave solutions, resulting in solutions of higher amplitude being relatively %less accurate than those of lower amplitude. Otherwise, the nonlocal  formula should never result in errors larger %than round-off due to machine precision, since the pressure at the bottom was computed using this nonlocal formula.

\begin{table}[tb]
\centering
\begin{tabular}{c|cccccc}
Percentage of & & &  &  &  & \\
Limiting Wave Height & SWO1 & SAO1 & SWO3 & SAO2 & SAO2h & Nonlocal \\
\hline
35 & 21.14 & 9.43 & 4.01 & 2.36 & 1.76 & 0.00 \\
45 & 25.67 & 13.43 & 6.49 & 4.27 & 3.18 & 0.00 \\
50 & 27.88 & 15.49 & 7.89 & 5.41 & 4.04 & 0.00 \\
53 & 28.94 & 16.49 & 8.67 & 5.96 & 4.40 & 0.00 \\
54 & 29.65 & 12.17 & 9.15 & 6.36 & 4.70 & 0.00 \\
55 & 30.08 & 17.58 & 9.45 & 6.61 & 4.88 & 0.00
\end{tabular}
\caption{Relative error \rf{relerror} in percent, calculated comparing peak wave heights using various reconstruction formulas using numerical data.}\label{moderate}
\end{table}

\begin{figure}[tb]
\centering \includegraphics[width=.95\textwidth]{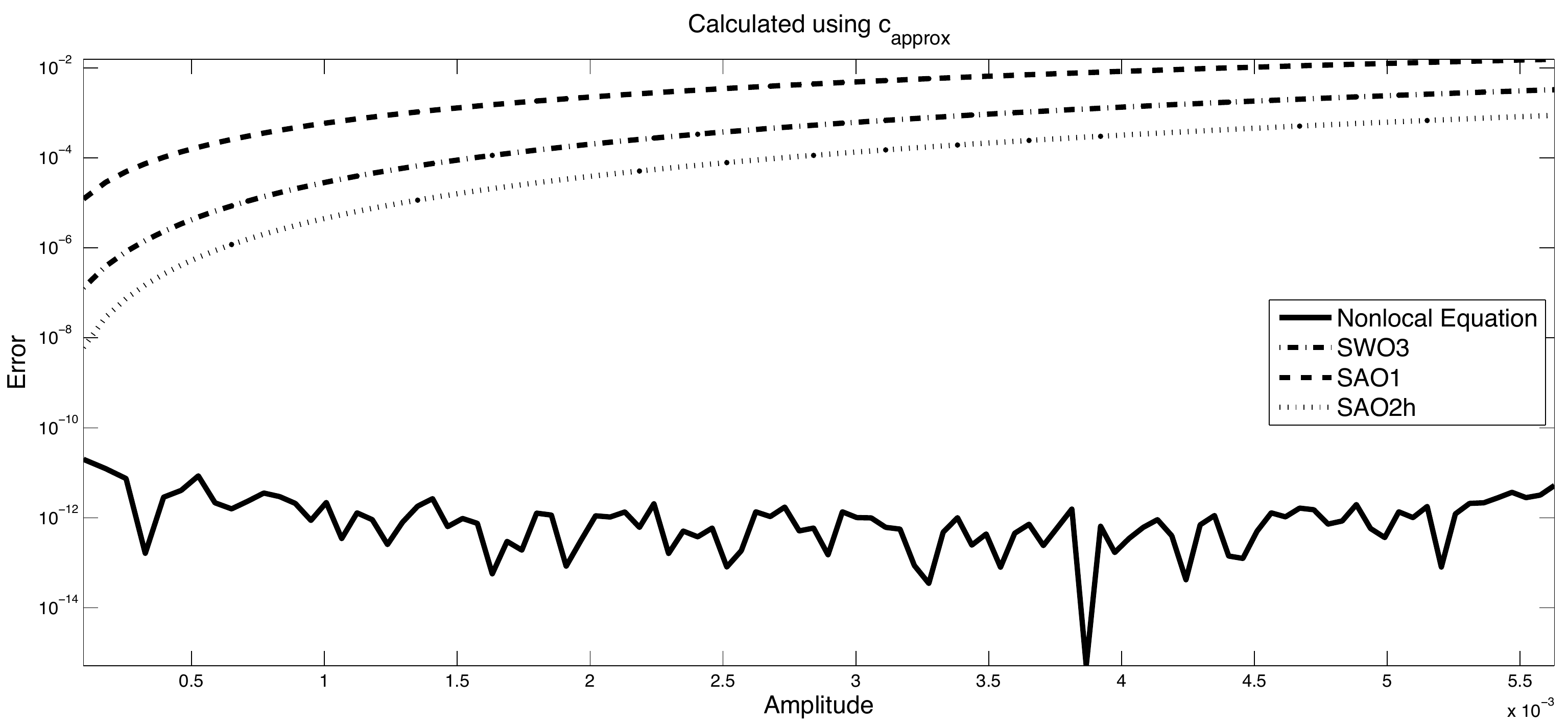}
{\caption{Plot of the error in the reconstructed surface elevation $\eta_r$ as a function of the amplitude of $\eta_{true}$ using an approximation for the wave-speed $c$. The SAO2 model is not included for the same reason as in Figure~\ref{fig:NumericErrorRealC}.}\label{fig:NumericErrorEstimateC}}
\end{figure}

One limitation of the nonlocal equation \rf{awesome} and some of its asymptotic counterparts from the previous section is that they require the knowledge of the traveling wave speed $c$.  In practice, this can be a difficult or impractical quantity to measure.  One such impractical option is to include additional pressure sensors in order to measure the time it takes for the peak of the pressure data to travel from one sensor to another.  A  simpler option is to use approximations for the wave speed based on small-amplitude theory.  For example, if we repeat the same error calculation as above, but with $c\approx \sqrt{gh}$, we obtain Figure~\ref{fig:NumericErrorEstimateC}. We might hope that the reconstruction of the surface elevation would not suffer much.  In fact, it appears unchanged.  As seen in Figure~\ref{fig:NumericErrorEstimateC}, the error in reconstructing the peak wave height does not suffer at all from using this simple approximate (and amplitude-independent) value of $c$.  Surprisingly, the error in the nonlocal reconstruction remains consistent with the error calculated using the true wave speed $c$.  This lack of sensitivity to the precise value of $c$ yields hope that with experimental data a simple approximation of the wave speed will be sufficient to accurately reconstruct the surface elevation.

\subsection{Comparison of the Different Approaches Using Data From Physical Experiments}

Here we discuss comparisons of results from the nonlocal formula ({\ref{awesome}}) and the asymptotic approaches with results from ten laboratory experiments performed at Penn State's Pritchard Fluid Mechanics Laboratory. In these experiments the pressure at the bottom of the fluid domain and the displacement of the air--water interface were measured simultaneously. The experimental facility consisted of the wave channel and water, the wavemaker, bottom pressure transducers, and a surface displacement measurement system. The wavetank is 50\,ft long, 10\,in wide and 1\,ft deep. It is constructed of tempered glass. It was filled with tap water to a depth of $h$, as listed in Table~\ref{tab:experimentalError}. The pressure gage was a SEN$^Z$ORS PL6T submersible level transducer with a range of 0--4\,in.
It provided a 0--5\,V dc output, which was digitized with an NI PCI-6229 analog-to-digital converter using LabView software. We calibrated this transducer by raising and lowering the water level in the channel. The pressure measurements had a high-frequency noise component, and thus were low-pass filtered at 20\,Hz.
The still-water height was measured with a Lory Type C point gage.
The capacitance-type surface wave gage consisted of a coated-wire probe connected to an oscillator. The difference frequency between this oscillator and a fixed oscillator was read by a Field Programmable Gate Array (FPGA), NI PCI-7833R. Thus, no D/A conversion, filtering, or A/D conversion was required. The surface capacitance gage was held in a rack on wheels that are attached to a programmable belt. We calibrated the capacitance gage by traversing the rack at a known speed over a precisely machined, trapezoidally-shaped ``speed bump''.
The waves were created with a horizontal, piston-like motion of a paddle made from a Teflon plate (0.5\,in thick) inserted in the channel cross-section. The paddle was machined to fit the channel precisely with a thin lip around its periphery that served as a wiper with the channel's glass perimeter. This wiper prevented any measurable leakage around the paddle during an experiment. The paddle was connected to the programmable belt and traveled in one direction. It was programmed with the horizontal velocity of a KdV soliton, which is given by

\beq
u(x,t) = u_0\, {\rm{sech}}^2\left( \frac{3 u_0}{4 {h_0}^2 c_0} (x - c_0 t - u_0 t/2) \right),
\label{wavemaker}
\eeq

\noindent where $c_0=\sqrt{g h}$, $a_0$ is the wave amplitude, and $u_0 = a_0 c_0/h$ is the maximum horizontal velocity. The $a_0$ for the wavemaker displacement was varied between 2\,cm and 3\,cm. These values corresponded to large velocities and fluid displacements, outside of the regime of the KdV equation. The water adjusted to create a leading wave with a radiative tail. We compare results for the leading wave, where nonlinearity is likely to be important.

%Here we use the time series of data for the pressure at the bottom of the tank and for the surface elevation directly above the
%pressure sensor to compare the results obtained using the different asymptotic approaches and those from the nonlocal
%formula \rf{awesome}.
To convert the time series of pressure and surface displacement data into spatial data, we use a combination of the sampling frequency and the estimated wave speed $c$. Specifically, let $p_j$ represents the measured pressure at time $t_j = j\,\Delta t$, where $\Delta t$ is the time between pressure measurements. We assign a corresponding $x$ value $x_j$ to $p_j$ so that $x_j = c \left(j\,\Delta t\right)$. From the pressure data measured from the physical experiments, we reconstruct the surface elevation using the same methods as in the previous section.
%\begin{figure}[tb]
%\centering \includegraphics[width=.9\textwidth]{./experiment_setup}
%{\caption{Time Series of pressure (top) and surface elevation (bottom).  The data have been %filtered}\label{fig:timeSeries_Data}}
%\end{figure}
%Starting with a time series of data for the pressure at the bottom of the tank, as well as for the surface elevation %directly above the pressure sensor, we wish to compare the results obtained using the different asymptotic approaches %and those from the nonlocal formula \rf{awesome}.  To convert the time series data into spatial data, we use a %combination of the sampling frequency and the estimated wave speed $c$. Specifically, let $p_j$ represents the %measured pressure at time $t_j = j\,\Delta t$, where $\Delta t$ is the time between pressure measurements. We assign a %corresponding $x$ value $x_j$ to $p_j$ so that $x_j = c \left(j\,\Delta t\right)$. From the pressure data measured %from the physical experiments, we reconstruct the surface elevation using the same methods as in the previous section, %with the additional step of prefiltering the data to remove high-frequency noise by passing through a low-pass filter %with a suitable cut-off frequency.
%A total of ten experiments are considered, using water of different depths and waves of different amplitudes. In these %experiments time series of the pressure at the bottom and of the surface elevation are measured directly. These data %are filtered and converted to spatial data as described above.
For all experiments we use the admittedly simple approximation $c \approx \sqrt{gh}$. We use the measured pressure data to reconstruct the surface elevation using the nonlocal formula \rf{awesome}, as well as the asymptotic approximations SWO1 (hydrostatic), SAO1 (transfer function), SAO2, and SWO3.

\begin{figure}[tb]
\centering \includegraphics[width=.95\textwidth]{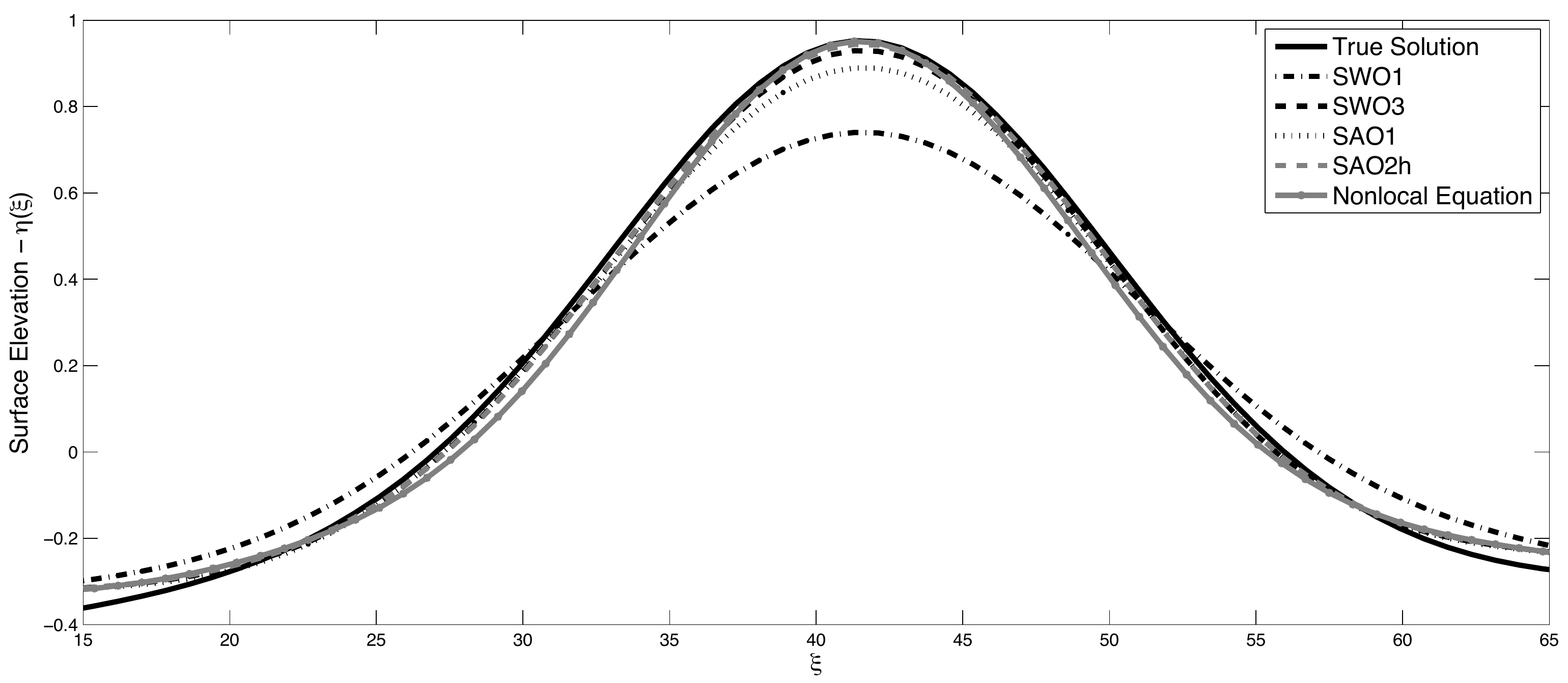}
\caption{Wave tank comparisons of the reconstructed surface elevation with surface height measurements for $h = 5.05\,cm$.  This corresponds to Experiment \#1 in Table~\ref{tab:experimentalError}.
\label{fig:ExperimentReconstruction_505_ex_1}}
\end{figure}

\begin{figure}[tb]
\centering \includegraphics[width=.9\textwidth]{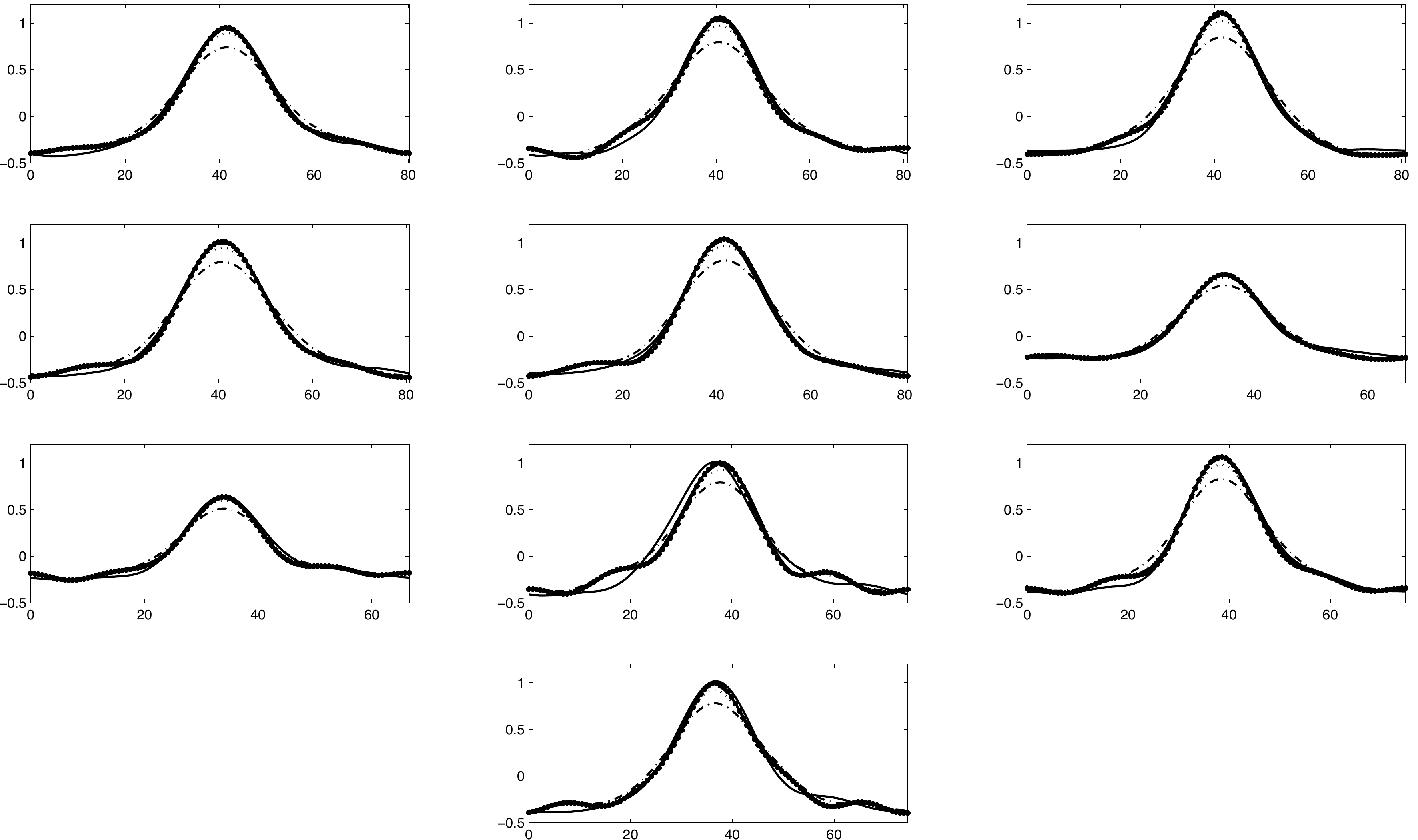}
{\caption{Wave tank comparisons of the reconstructed surface elevation with surface height measurements for various fluid depths. The experiments are ordered from left-to-right and top-to-bottom, corresponding to the experiment \# in Table~\ref{tab:experimentalError}.}\label{fig:ExperimentReconstruction_Multiple}}
\end{figure}

As seen in Figure~\ref{fig:ExperimentReconstruction_505_ex_1}, the higher-order methods capture the peak wave height better than the lower-order methods, with the nonlocal equation yielding the most accurate representation of the peak wave height. The visual comparisons for all experiments is displayed in Figure~\ref{fig:ExperimentReconstruction_Multiple}. The nonlocal formula \rf{awesome} consistently captures the peak wave height better than any of the approximate models derived in the previous section, and significantly better than the SWO1 (hydrostatic) and SAO1 (transfer function) models. This is quantified in Table~\ref{tab:experimentalError}, which displays the error \rf{relerror} for the different approximations and the nonlocal equation \rf{awesome}. As is seen there, the result from \rf{awesome} consistently produced the smallest error among all the reconstruction formulas.  It is noteworthy that the heuristic SAO2h approximation \rf{secret} consistently yields the second-lowest peak height error and consistently outperforms all other models except the nonlocal equation \rf{awesome}. Given the computational expense of solving the nonlocal equation, the approximation SAO2h apparently yields the best compromise between efficiency and accuracy.

\begin{table}[tb]
\centering
\begin{tabular}{c|c||c|c|c|c|c|c}
Experiment \# & Depth (cm) & SWO1~ & SAO1~ & SWO3~ & SAO2~ & SAO2h & Nonlocal\\
\hline\hline
1 & 5.05  & 22.29 & 6.51 & 2.35 & 0.84 & 0.45 & 0.20 \\
2 & 5.05  & 24.66 & 8.05 & 3.30 & 1.13 & 0.56 & 0.01 \\
3 & 5.05  & 23.56 & 7.75 & 2.90 & 0.95 & 0.41 & 0.18 \\
4 & 5.05  & 21.98 & 7.18 & 2.80 & 1.29 & 0.89 & 0.66 \\
5 & 5.05  & 22.00 & 6.63 & 2.10 & 0.48 & 0.05 & 0.03 \\
6 & 3.55  & 18.11 & 5.21 & 1.65 & 0.65 & 0.43 & 0.36 \\
7 & 3.55  & 20.81 & 7.04 & 3.33 & 2.22 & 1.95 & 1.52 \\
8 & 4.10  & 21.59 & 8.39 & 3.50 & 2.18 &1.73 & 0.93 \\
9 & 4.10  & 22.13 & 7.65 & 2.29 & 0.33 & 0.31 & 0.05 \\
10 & 4.10 & 23.32 & 9.25 & 4.60 &2.91 & 2.44 & 2.13
\end{tabular}
\caption{Relative error \rf{relerror} in percent, calculated comparing peak wave heights using various reconstruction formulas using experimental data.}\label{tab:experimentalError}
\end{table}

\section{Conclusion}

We have presented a new equation \rf{awesome} relating the pressure at the bottom of the fluid to the surface elevation of a traveling wave solution of the one-dimensional Euler equations without approximation. This equation is analyzed rigorously and the existence of solutions is proven using the Implicit Function Theorem. Solving the equation numerically is possible, but this is computationally relatively expensive when compared to currently-used approaches that require the computation of at most a few Fourier transforms. To this end, we derive various new approximate formulas, starting from the new nonlocal formula. The canonical approaches (hydrostatic approximation and transfer function approach) are easily obtained from the nonlocal formula as well.

The different approximations and the nonlocal formula \rf{awesome} are compared using numerical data, and their performance on physical laboratory data is examined. The nonlocal formula consistently outperforms its different approximations. For the numerical data this is by construction, as it was used to generate the numerical pressure data used for the comparison, starting from computed traveling wave solutions of the Euler equation. The higher-order approximate formulas result in a better reconstruction of the surface elevation compared to the hydrostatic or transfer function approaches. In the lab experiments, both the surface elevation and the bottom pressure are measured, allowing for an independent validation of the nonlocal equation \rf{awesome}. As expected, it outperforms the different approximations, where higher-order models perform better than lower-order ones. A good compromise between computational cost and obtained accuracy seems to be achieved by the heuristic approximation SAO2h \rf{secret}, which requires the computation of three Fourier transforms.

Our derivation of the nonlocal equation \rf{awesome} requires the surface elevation profile to be traveling with constant speed $c$. Regardless, we show that the results are not sensitive to the exact value of $c$ and even rough estimates ({\em i.e.}, $c=\sqrt{gh}$) provide excellent results, both for the nonlocal equation \rf{awesome} and its various asymptotic approximations, most notably \rf{secret}.

\section*{Acknowledgements}

BD and VV acknowledge support from the National Science Foundation under grant NSF-DMS-1008001. DH acknowledges support from the National Science Foundation under grants NSF-DMS-0708352 and NSF-DMS-1107379. She is grateful to Rod Kreuter and Rob Geist for development of the electrical and mechanical systems for the experiments. Any opinions, findings, and conclusions or recommendations expressed in this material are those of the authors and do not necessarily reflect the views of the funding sources.

\end{document}